\documentclass[twocolumn,english,aps,prb,twocolum,superscriptaddress,bibnotes,amsmath,amssymb,floatfix]{revtex4-1}
\usepackage[colorlinks=true,citecolor=blue,linkcolor=magenta]{hyperref}

\usepackage[utf8]{inputenc}
\usepackage[english]{babel}
\usepackage{amsmath,amsfonts,amssymb}
\usepackage[T1]{fontenc}
\usepackage{url}
\usepackage{soul}

\usepackage{amsmath}
\usepackage{amsfonts}
\usepackage{amssymb}
\usepackage[version=3]{mhchem}
\usepackage{stmaryrd}
\usepackage{epstopdf}
\usepackage{graphicx}
\graphicspath{{./Figures/}}

\begin{document}

\title{Electrically pumped photonic integrated soliton microcomb}

\author{Arslan S. Raja}
\thanks{These authors contributed equally to the work}
\affiliation{{\'E}cole Polytechnique F{\'e}d{\'e}rale de Lausanne (EPFL), CH-1015 Lausanne, Switzerland}
 
\author{Andrey S. Voloshin}
\thanks{These authors contributed equally to the work}
\affiliation{Russian Quantum Center, Moscow, 143025, Russia}

\author{Hairun Guo}
\thanks{These authors contributed equally to the work}
\affiliation{{\'E}cole Polytechnique F{\'e}d{\'e}rale de Lausanne (EPFL), CH-1015 Lausanne,
Switzerland}

\author{Sofya E. Agafonova}
\thanks{These authors contributed equally to the work}
\affiliation{Russian Quantum Center, Moscow, 143025, Russia}
\affiliation{Moscow Institute of Physics and Technology, Dolgoprudny, 141700, Russia}

\author{Junqiu Liu}
\thanks{These authors contributed equally to the work}
\affiliation{{\'E}cole Polytechnique F{\'e}d{\'e}rale de Lausanne (EPFL), CH-1015 Lausanne,
Switzerland}

\author{Alexander S. Gorodnitskiy}
\affiliation{Russian Quantum Center, Moscow, 143025, Russia}
\affiliation{Moscow Institute of Physics and Technology, Dolgoprudny, 141700, Russia}

\author{Maxim Karpov}
\affiliation{{\'E}cole Polytechnique F{\'e}d{\'e}rale de Lausanne (EPFL), CH-1015 Lausanne,
Switzerland}

\author{Nikolay G. Pavlov}
\affiliation{Russian Quantum Center, Moscow, 143025, Russia}
\affiliation{Moscow Institute of Physics and Technology, Dolgoprudny, 141700, Russia}

\author{Erwan Lucas}
\affiliation{{\'E}cole Polytechnique F{\'e}d{\'e}rale de Lausanne (EPFL), CH-1015 Lausanne,
Switzerland}

\author{Ramzil R. Galiev}
\affiliation{Russian Quantum Center, Moscow, 143025, Russia}
\affiliation{Faculty of Physics, M.V. Lomonosov Moscow State University, 119991 Moscow, Russia}

\author{Artem E. Shitikov}
\affiliation{Russian Quantum Center, Moscow, 143025, Russia}
\affiliation{Faculty of Physics, M.V. Lomonosov Moscow State University, 119991 Moscow, Russia}

\author{John D. Jost}
\affiliation{{\'E}cole Polytechnique F{\'e}d{\'e}rale de Lausanne (EPFL), CH-1015 Lausanne,
Switzerland}

\author{Michael L. Gorodetsky}
\email[]{mg@rqc.ru}
\affiliation{Russian Quantum Center, Moscow, 143025, Russia}
\affiliation{Faculty of Physics, M.V. Lomonosov Moscow State University, 119991 Moscow, Russia}

\author{Tobias J. Kippenberg}
\email[]{tobias.kippenberg@epfl.ch}
\affiliation{{\'E}cole Polytechnique F{\'e}d{\'e}rale de Lausanne (EPFL), CH-1015 Lausanne,
Switzerland}

\date{\today}

\maketitle

\noindent\textbf{\noindent
Optical frequency combs have revolutionized frequency metrology and timekeeping, and can be used in a wide range of optical technologies. Advances are under way that allow dramatic miniaturization of optical frequency combs using Kerr nonlinear optical \linebreak microresonators, where broadband and coherent optical frequency combs can be generated from a continuous wave laser. Such `microcombs', provide a broad bandwidth with low power consumption, unprecedented form factor, wafer scale fabrication compatibility, and can potentially allow combs to be field deployed, outside of research laboratories. For future high volume applications, integration and electrical pumping of soliton microcombs is essential. To date, however, microcombs still rely on optical pumping by bulk external laser modules that provide the required coherence, frequency agility and power levels for soliton formation.  Electrically-driven, chip-integrated microcombs are inhibited by the high threshold power for soliton formation, typically exceeding the power of integrated silicon based lasers, and the required frequency agility for soliton initiation. Recent advances in high-$\mathbf{Q}$ $\mathbf{Si_3N_4}$ microresonators  suggest that electrically driven soliton microcombs are possible. Here we demonstrate an electrically-driven, chip-integrated soliton microcomb by coupling an indium phosphide (III-V) multiple-longitudinal-mode laser diode chip to a high-$\mathbf{Q}$ $\mathbf{Si_3N_4}$ photonic integrated microresonator. We observe that self-injection-locking of the laser diode to the microresonator, which is accompanied by a $\mathbf {\times1000}$ fold narrowing of the laser linewidth, can simultaneously initiate  the formation of dissipative Kerr solitons.
By tuning the current, we observe a transition from modulation instability, breather solitons to single soliton states. The resulting soliton self-injection-locking-based microcomb exhibits narrow comb-teeth linewidth ($\mathbf{\sim 100}  \mathbf{kHz}$). The system requires less than 1 $ \mathbf{Watt}$  of electrical power, operates at electronically-detectable sub-100 $\mathbf{GHz}$ mode spacing and can fit in a volume of ca. $\mathbf{1cm^3}$. This approach, compatible with semiconductor laser diode technology, provides a route towards scalable manufacturing of microcombs as required for high-volume applications such as laser ranging or optical interconnects.
}

\begin{figure*}[t!]
  \centering{
  \includegraphics[width = 0.85 \linewidth]{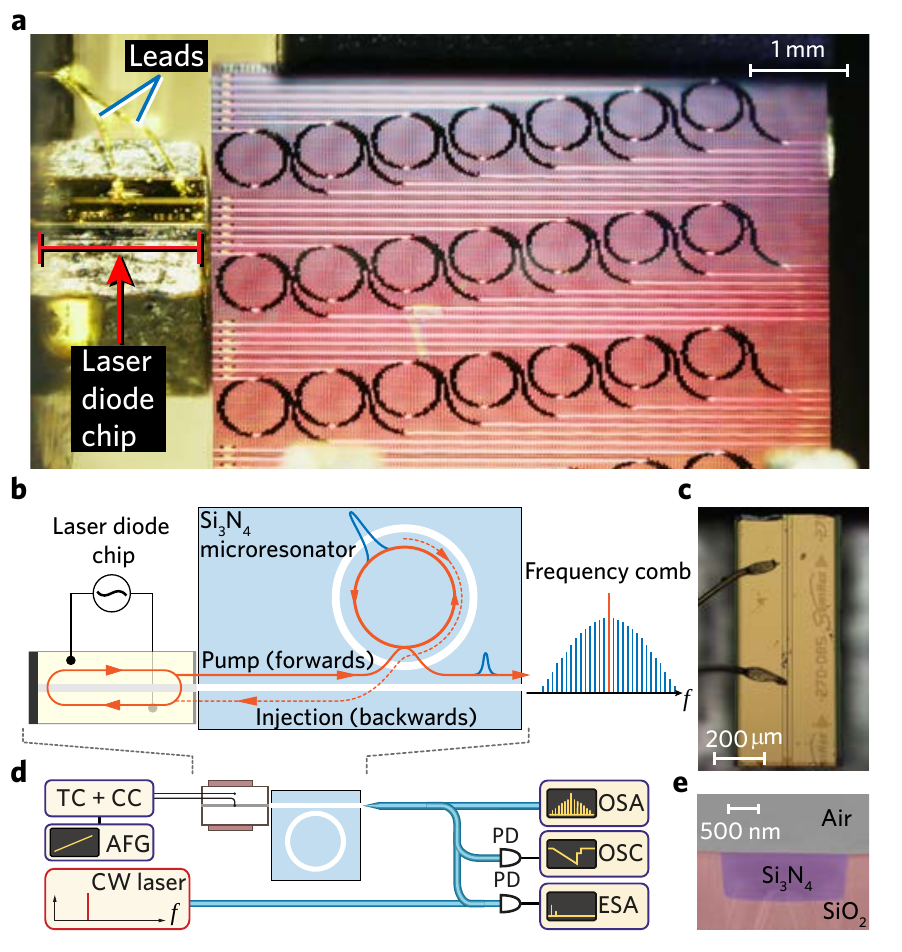}
  }
  \caption{ \noindent \textbf{Principle of an ultra-compact, laser-injection-locked soliton Kerr frequency comb.}
 \textbf{(a)} Close-range photo of the experimental setup, in which the laser diode chip is butt-coupled to a ${\rm{Si_3N_4}}$ photonic chip, which contains several microresonators.  \textbf{(b)} Schematic representation of the laser-injection-locked soliton Kerr frequency comb. An InP  multi-frequency laser diode chip is directly butt-coupled to a ${\rm Si_3N_4}$ photonic chip with a microresonator.  \textbf{(c)} An optical image  of the InP  laser diode chip showing the magnified view. \textbf{(d)} Sketch of the experimental setup. The microresonator device output is characterized both in the optical domain using an optical spectral analyzer, and in the radio frequency (RF) domain using an electrical-signal spectral analyzer. In addition, to assess the coherence of the frequency comb, we employ a heterodyne beatnote measurement to a selected comb tooth with a narrow-linewidth reference laser. TC: temperature control module. CC: current control module. AFG: arbitrary function generator.  OSA: optical spectral analyzer. OSC: oscilloscope. ESA: electrical-signal spectral analyzer. \textbf{(e)} False-colored scanning electron micrograph (SEM) image of the waveguide cross section. The ${\rm Si_3N_4}$ waveguide (blue shaded) has no top SiO$_2$ cladding but only side and bottom SiO$_2$ cladding (red shaded). 
  \label{fig_setup}
}
\end{figure*} 
Optical frequency combs \cite{Udem2002metrology,Cundiff2003comb} have revolutionized time-keeping and frequency metrology over the past two decades, and have found a wide variety of applications.
Microresonator-based Kerr frequency combs (Kerr microcombs) \cite{del2007optical,kippenberg2011microresonator} have provided a route to compact chip-scale optical frequency combs, with broad optical bandwidth and repetition rates in the microwave to terahertz domain (${10~{\rm GHz}-1~{\rm THz}}$). Their compact and low-power nature could enable utilization in mobile or airborne applications beyond research laboratories, including operation in space\cite{brasch2014radiation}. The observation that such microcombs can be operated in the dissipative Kerr soliton (DKS) regime (soliton microcombs) \cite{leo2010temporal,herr2014temporal,brasch2016photonic}, 
has enabled fully coherent microcombs \cite{herr2014temporal}. DKS exhibit a rich set of nonlinear optical phenomena such as soliton Cherenkov radiation (also known as dispersive waves) which can extend the spectral bandwidth  of the frequency comb \cite{ brasch2016photonic}.
Soliton microcombs have been applied in counting of the cycles of light \cite{jost2015counting}, coherent communication \cite{marinpalomo2017communication}, ultrafast ranging \cite{trocha2017ranging,suh2018ranging}, dual-comb spectroscopy \cite{suh2016dualcomb}, low-noise microwave generation \cite{liang2015microwave}
and optical frequency synthesis\cite{spencer2018optical}.
The full photonic integration of soliton microcombs in a single, compact, and electrically-driven package that use compact chipscale lasers would allow mass-manufacturable devices compatible with emerging high-volume applications such as laser-based ranging (LIDAR), or sources for dense wavelength division multiplexing for data center-based optical interconnects. Via advances in silicon photonics, such level of integration has been achieved for lasers\cite{fang2006electrically}, modulators \cite{green2007siliconmodulator}, and a wide range of passive and active elements which are already commercially available.
Photonic integration of soliton microcombs requires not only the integration of nonlinear high-${Q}$ microresonators on chip, but also an on-chip solution for the narrow linewidth seed lasers with output power levels that are sufficient for soliton initiation,  as well as any laser tuning mechanism\cite{herr2014temporal,stone2017thermal,brasch2016photonic,brasch2016bringing}  used in the soliton excitation process. 
Photonic integration of high-${Q}$ microresonators suitable for soliton formation has advanced significantly, in particular using ${\rm Si_3N_4}$ -- a CMOS-compatible material used as a capping layer. The platform possesses several advantageous properties \cite{levy10cmossin}, including a high Kerr nonlinearity, large flexibility for dispersion engineering, outer-space compatibility  \cite{brasch2014radiation}, and a large bandgap $( \mathrm {\sim 5eV)}$, thus free from two-photon absorption in the telecommunication band. All these advantages facilitate soliton formation in ${\rm Si_3N_4}$ microresonators \cite{brasch2016photonic}. In a related effort, ultrahigh-${Q}$ $\mathrm{SiO}_2$ air-clad microresonators have recently been integrated with ${\rm Si_3N_4}$ waveguides for soliton generation \cite{yang2018bridging}. Efforts to combine  integrated photonic microresonators with chip-scale lasers, such as those developed in silicon photonics, have recently been made\cite{spencer2018optical}. 
Yet, these and other approaches still rely on stand-alone bulk laser modules,  and typically employ additional amplifiers for soliton initiation to overcome coupling losses and the low $Q$-factors of integrated photonic resonators. Likewise, the use of silicon photonics-based lasers is presently compounded by the threshold of soliton formation that typically significantly exceeds the laser's output power (mW scale). 
Recent advances in fabrication of high-$Q$  ${\rm Si_3N_4}$  photonic integrated microresonators (intrinsic ${Q_0 >  1 \times 10^7}$) \cite{xuan2016high-q,ji2017submilli,pfeiffer2018probing} suggest that electrically driven microcombs that employ chip-scale laser diodes -- compatible with scalable manufacturing -- may become viable.

Here we demonstrate an electrically-driven, and current-initiated, soliton microcomb significantly simplifying photonic integration \cite{patent2018injection}. The integrated device has a volume of ca. $1\mathrm{cm^3}$,  uses a commercially available semiconductor laser diode chip. This device consumes less than 1 Watt of electrical power and produces soliton microcomb with sub-100-GHz line spacing. By using high-${Q}$ (${Q_0> 1 \times 10^7}$) photonic chip-scale ${\rm Si_3N_4}$ microresonators  fabricated using the photonic Damascene reflow process \cite{pfeiffer2018probing,liu2018ultralow-power}, in conjunction with a multiple-longitudinal-mode (multi-frequency)
Fabry-P{\'e}rot InP laser diode chip, we observe self-injection locking \cite{dahmani1987fpinjection,vassiliev1998narrowinjection} in a regime where solitons are formed concurrently. Such self-injection locking with concurrent soliton formation has recently been demonstrated for bulk ultrahigh-${Q}$ crystalline $\mathrm{MgF_2}$ resonators \cite{liang2015microwave,Pavlov2018NP}. 
We observe that the current tuning of the laser diode can induce transitions from the injection-locking-based single-longitudinal-mode lasing ($\times 1000$ fold reduction of linewidth), to Kerr frequency combs, breather soliton formation, followed by stable multiple and single DKS formation in the integrated microresonator. Heterodyne measurements demonstrate the low-noise nature of the generated soliton states.  Such an electrically-driven photonic chip-based soliton microcomb demonstrated here, provides a solution for integrated, unprecedentedly compact optical comb sources suitable for high volume applications. We note that concurrent to this report, in a related approach, another integrated soliton microcomb has been demonstrated using a semiconductor optical amplifier coupled to a high-${Q}$ ${\rm Si_3N_4}$ microresonator\cite{stern2018fully}.

Figure \ref{fig_setup} illustrates the approach taken in this work.
A  multi-frequency Fabry-P{\'e}rot laser diode chip (InP) is directly butt-coupled to a ${\rm Si_3N_4}$ photonic chip [Fig. \ref{fig_setup}(a,b)]. The butt-coupling scheme gives an overall insertion loss of ${\sim 6 ~{\rm dB}}$ (diode-chip-lensed fiber), with a double inverse tapered structure for the light input/output coupling\cite{liu2018doubleinverse}. 
When the frequency of the light emitted from the laser diode coincides with a high-${Q}$ resonance of the Si$_3$N$_4$ microresonator, laser self-injection locking can take place.
The process occurs due to the bulk and surface Rayleigh scattering in the microresonator, which injects a fraction of light back into the diode \cite{vassiliev1998narrowinjection}. This provides a frequency-selective optical feedback to the laser, leading to single-frequency operation and a significant reduction of the laser linewidth.

A key step for our approach is to match the optical power requirement for soliton generation to that of the laser diode. This is achieved by employing high-${Q}$ ${\rm Si_3N_4}$ microresonators fabricated using the photonic Damascene process, featured with a novel and crucial reflow step \cite{pfeiffer2018probing}, allowing for ultra-smooth waveguide sidewalls that enable high-${Q}$ factors (${Q_0 > 1 \times 10^7}$) across the entire L band (cf. Methods).

\begin{figure*}[t!]
  \centering{
 \includegraphics[width = 1 \linewidth]{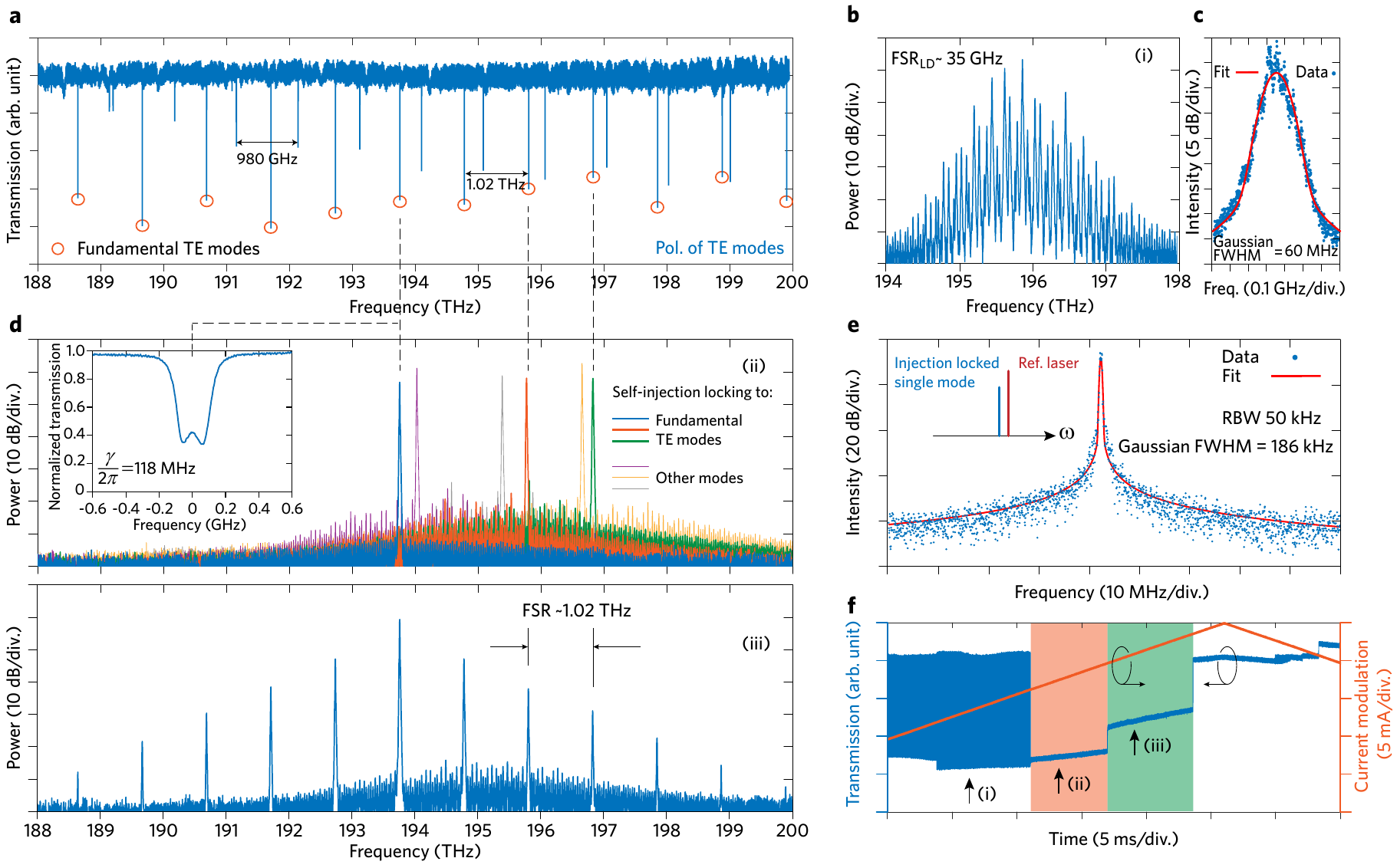}
  }
  \caption{ \noindent  \textbf{Electrically pumped soliton microcomb via laser-injection-locked soliton formation.}
   \textbf{(a)}  Transmission spectrum of a ${\rm{Si_3N_4}}$ microresonator of 1.02 THz FSR, featuring two sets of resonances: the fundamental transverse electric (TE) mode family (marked by red circles) and one high-order TE mode family. \textbf{(b)}  The laser spectrum of the multi-frequency laser diode chip used in this experiment, corresponding to state i in (f). \textbf{(c)}  Measured and fitted heterodyne beat signal between the free running  laser diode and a narrow-linewidth reference laser (Toptica CTL1550, short-time linewidth ${\sim 10 ~{\rm kHz}}$), showing 60 MHz linewidth. \textbf{(d)}  Top panel (state ii in \textbf{(f)}): Spectra of single-longitudinal-mode that is injection-locked to a selected resonance of the microresonator.  Bottom panel (state iii in \textbf{(f)}): Spectrum of the Kerr frequency comb that stems from the laser injection locking. Inset: One resonance of the fundamental TE mode showing mode splitting due to backscattering, with the estimated ${118 ~{\rm MHz}}$ coupling strength between the forward and backward propagating modes. \textbf{(e)}  Heterodyne beat signal between the injection-locked  laser and a narrow-linewidth reference laser. The measured beat signal is fitted with Voigt profile with full width at half maximum (FWHM) ${ \sim 186 ~{\rm kHz}}$ (cf. Methods). \textbf{(f)}  Typical transmitted power trace measured at the chip output facet, by current modulation imposed on the laser diode, in which different states are marked: (i) Noisy, multi-frequency  lasing without injection locking; (ii) Laser injection locking to a microresonator resonance, and simultaneous formation of low-noise single-longitudinal-mode lasing; (iii) Formation of Kerr frequency comb.
  \label{fig_kickOn}
}
\end{figure*}

The Fabry-P{\'e}rot laser diode we employ in the experiments is centered at ${1530 ~{\rm nm}}$, and its emission spectrum without self-injection locking 
 is shown in Fig. \ref{fig_kickOn}(b). The mode spacing is ${35 ~{\rm GHz}}$, determined by the Fabry-P{\'e}rot cavity length. The overall maximum optical output power is ${\sim 100 ~{\rm mW}}$ when applying a current of ${\sim 350 ~{\rm mA}}$ to the diode.
The electrical power consumed by the laser diode is  less than ${1 ~{\rm W}}$. Figure \ref{fig_kickOn}(c) shows the heterodyne beatnote of the free running laser diode mode with the reference laser (Toptica CTL1550, short-time linewidth ${\sim 10 ~{\rm kHz}}$), which is fitted with the Voigt profile (cf. discussions in the following and Methods), revealing both a Gaussian  linewidth of ${60 ~{\rm MHz}}$ and an estimated short-time linewidth of  ${2 ~{\rm MHz}}$.

We first studied self-injection locking of the laser diode chip to the photonic chip-based microresonator.  This is achieved by tuning the current of the laser diode, which not only changes the optical output power, but also shifts the lasing frequency via the carrier dispersion effect. 
Initially, the laser diode coupled to the ${\rm{Si_3N_4}}$ chip operates multi-frequency [Fig. \ref{fig_kickOn}(b)], a regime where none of the high-$Q$ microresonator modes is frequency-matched with the multi-mode laser emission of the diode. 
By shifting the lasing frequency of the diode via current tuning, we observe that the initially multi-frequency emission spectrum switches to single mode operation, indicative of self-injection locking. Fig. \ref{fig_kickOn}(d) demonstrates that the lasing frequency coincides with a selected resonance of the microresonator, and we also observe injection locking occurring for several resonances.
We note that all resonances, which give rise to the laser self-injection locking, feature mode splitting as a result of backscattering [cf. the inset in Fig. \ref{fig_kickOn}(d)]. 
The back-coupling rate for the measured resonance, extracted from its mode-splitting profile, is $\gamma/2\pi= {118 ~{\rm MHz}}$ (cf. Methods). 
The presence of this back-coupling leads to an amplitude reflection coefficient ($r$) from the passive microresonator on resonance:
\begin{equation}
r\approx\frac{2\eta\Gamma}{1+\Gamma^2},
\end{equation}
where $\eta={\kappa_{\mathrm{ex}}}/{\kappa}$ characterizes coupling efficiency ($\kappa = \kappa_{\mathrm{0}}+\kappa_{\mathrm{ex}} $, with $\eta=1/2$ corresponding to critical coupling, and $\eta\approx 1$ corresponding to strong overcoupling), and $\Gamma = {\gamma}/{\kappa}$ is the normalized mode-coupling parameter that describes the visibility of the resonance split.  
 According to Ref. \cite{kondratiev2017selfinjection} this reflection can initiate self-injection locking, and give rise to a narrow linewidth of:
\begin{equation}
\delta\omega\approx \delta\omega_{\rm free} \frac{Q_{\mathrm{LD}}^2}{Q^2}\frac{1}{16r^2(1+\alpha_g^2)},
\end{equation}
where $Q={\omega}/{\kappa}$ is the microresonator quality factor, $\omega / 2 \pi$ is the light frequency, $\delta\omega_{\rm free} / 2 \pi$ is the linewidth of the free running laser. The phase-amplitude coupling factor $\alpha_g$ is the linewidth enhancement factor, given by the ratio of the variation of the real refractive index to the imaginary refractive index of the laser diode active region in response to a carrier density fluctuation \cite{henry1982theory} and takes typical values from 1.6 to 7. The InGaAsP/InP multiple quantum well laser diode has $\alpha_g = 2.5$. The laser diode quality factor {$Q_{\mathrm{LD}}$} can be estimated as $Q_{\mathrm{LD}}\approx\frac{\omega\tau_d R_o}{1-R_o^2},
\label{eq:narlw}$
where  $R_o$ is the amplitude reflection coefficient of the output laser mirrors, and $\tau_d$ is the laser cavity round trip. The reflection coefficient is a parameter of the laser diode and is given by the laser diode manufacturer as $R_o = \sqrt{0.05}$ as well as $\alpha_g$ = 2.5. Other experimentally determined parameters are $\kappa/2\pi \approx 110$ MHz, $\gamma/2\pi \approx118 $ MHz, $\eta \approx0.64$, $\Gamma \approx 1$, and $\tau_d={1}/{\rm FSR_{\rm diode}}=1/(35\,\rm GHz)= 28.6$ ps. The theoretical estimation for the narrowed linewidth is $\delta\omega/2\pi \sim 0.1$ kHz. 
We next compare these theoretical estimates of the self-injectioned locked linewidth to experiments.
The linewidth of the self-injection-locked single-longitudinal-mode laser is measured by a heterodyne measurement [see Fig. \ref{fig_kickOn}(e)]. The  lineshape is fitted with a Voigt profile, which represents a convolution of Lorentzian and Gaussian lineshape (cf. Methods), yielding a Gaussian contribution to the linewidth of ${186 ~{\rm kHz}}$. The estimated Lorentzian contribution amounts to ${0.7 ~{\rm kHz}}$, describing the wings of the measured beatnote.
Self-injection locking leads to a narrowing of the white noise of the laser diode \cite{kondratiev2017selfinjection}. Therefore, this value should be compared with the Lorentzian contribution in the Voigt profile (i.e. ${0.7 ~{\rm kHz}}$) corresponding to a more than 1000-fold reduction in the linewidth. 
\begin{figure}[t!]
  \centering{
  \includegraphics[width = 1 \linewidth]{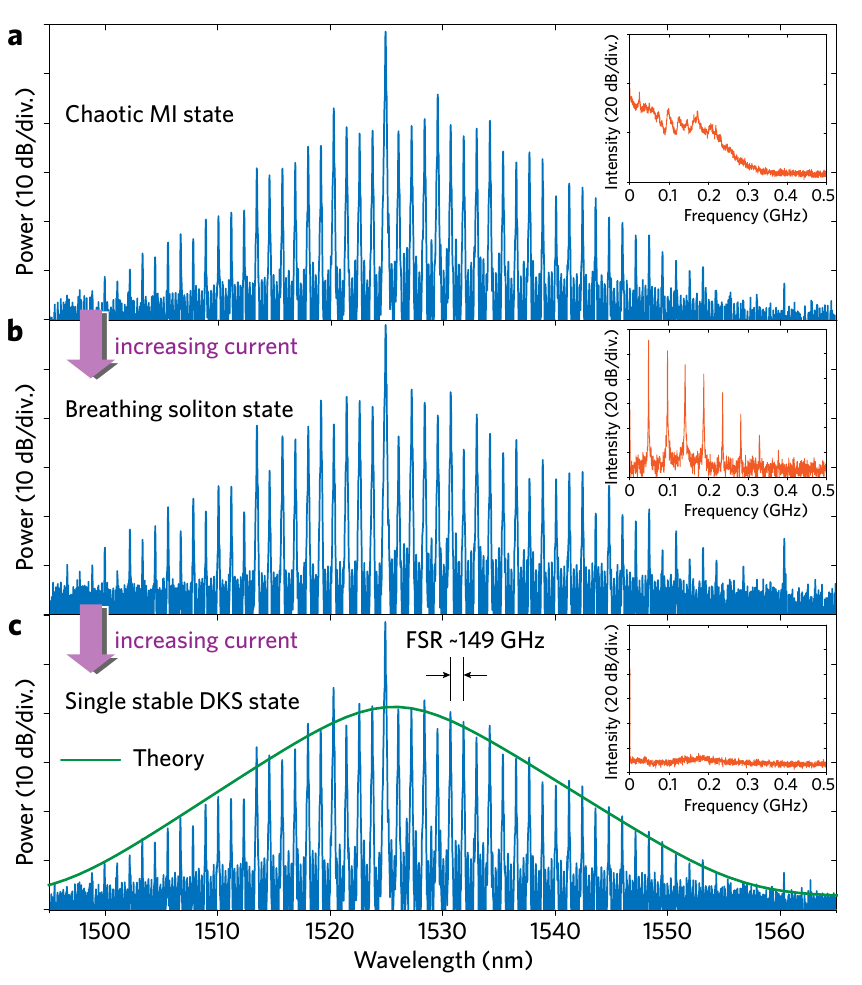}
  }
  \caption{ \noindent \textbf{Soliton comb generation with self-injection locking.}
   Evolution of Kerr frequency comb in the regime of laser self-injection locking, from noisy state in the operation regime of modulation instability \textbf{(a)} to breathing state \textbf{(b)}, and eventually to a low-noise state \textbf{(c)} showing the formation of a DKS in the microresonator, where the spectrum is a hyperbolic secant envelope (green-solid line showing the fitting of the spectral envelope). Each inset shows the low-frequency radio frequency (RF) spectrum corresponding to each state. The current imposed to the diode is initially set ${\sim 300 ~{\rm mA}}$ and the increase to evoke the transitions is within ${1 ~{\rm mA}}$. The ${\rm{Si_3N_4}}$ microresonator in this measurement has an FSR of ${149 ~{\rm GHz}}$.
  \label{fig_switch}
  }
\end{figure}

Injection locking occurs also in the case where the laser cavity and microresonator are detuned from each other due to ``injection pulling'',  and as outlined below, is imperative to generate DKS using self-injection locking. Injection pulling is a result of slight phase difference between the laser emission and its feedback, leading to imperfect locking \cite{kondratiev2017selfinjection}. The locking range is defined as the frequency range over which the laser diode emission self-injection locks to the high-$Q$ microresonator resonance
and follows the expression \cite{kondratiev2017selfinjection}:
\begin{equation}
\Delta\omega_{\rm lock}\approx r\sqrt{1+\alpha_g^2}\frac{\omega}{Q_{\mathrm{LD}}}.
\end{equation}
The theoretically estimated locking range exceeds $\Delta\omega_{\rm lock}/2\pi \approx 30$ GHz. 

Experimentally, we can access this effect by tuning the current of the laser diode, allowing the laser frequency to be changed concurrently with the self-injection locking, providing thereby a frequency scan over the resonance -- a prerequisite for DKS formation \cite{herr2014temporal}. 
Figure  \ref{fig_kickOn}(f) shows the optical output power (transmission) trace as a function of the current tuning, where self-injection locking is deterministically observed. 
An initial chaotic power trace [state (i) in Fig. \ref{fig_kickOn}(f)] is switched to a step-like pattern [state (ii) in Fig. \ref{fig_kickOn}(f), the orange marked region]. 
The average output power reduces during the switching since the self-injection leads to single-longitudinal-mode operation, with enhanced power being coupled into the high-${Q}$ resonance of the ${\rm Si_3N_4}$ microresonator. Most significantly, upon further tuning the current, a second step-like pattern in the power trace is observed [state (iii) in Fig. \ref{fig_kickOn}(f), the green marked region], corresponding to the formation of a (low noise) Kerr frequency comb.
Indeed, at high optical power levels (typically setting the current to be ${\sim 300 ~{\rm mA}}$), Kerr comb generation was observed upon tuning the current, as shown in Fig. \ref{fig_kickOn}(d).
This phenomenon relies critically on the ${Q}$-factor of the ${\rm Si_3N_4}$ microresonator, allowing sub-${{\rm mW}}$ threshold power for parametric oscillations (cf. Methods).  

\begin{figure*}[t!]
  \centering{
 \includegraphics[width = 1.0 \linewidth]{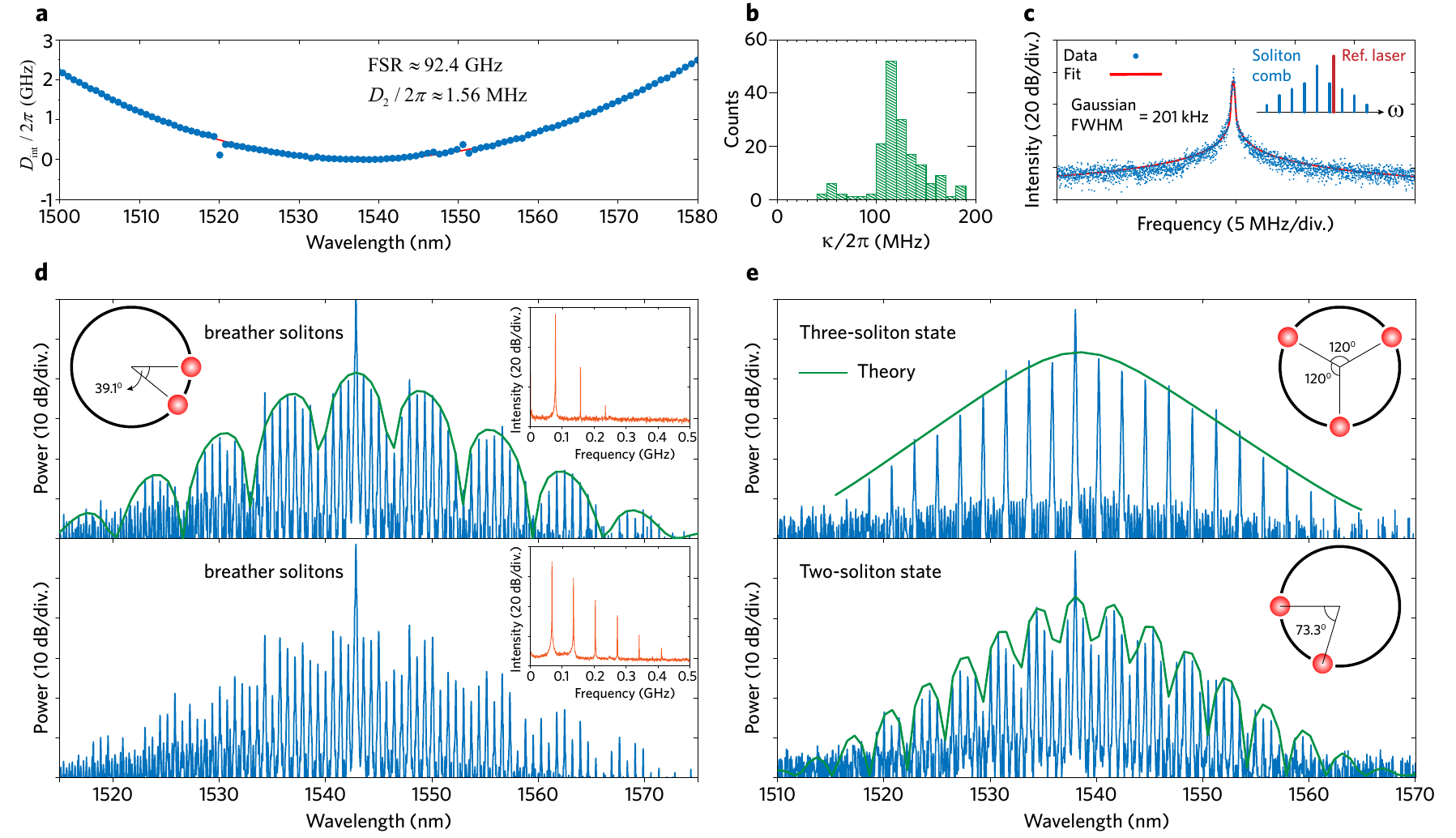}
  }
  \caption{ \noindent \textbf{Laser injection-locked multiple breathing and dissipative Kerr solitons. }
  \textbf{(a)} Measured and fitted dispersion landscape in a ${\rm{Si_3N_4}}$ microresonator (cross-section ${1.58 \times 0.75 ~{\rm \mu m^2}}$) (cf. Methods), which has the FSR ${=92.4 ~{\rm GHz}}$, and the second order dispersion element indicating the anomalous group velocity, ${D_2 / {2 \pi} \approx 1.56 ~{\rm MHz}}$. \textbf{(b)} Histogram of resonance linewidths that are ${\sim 110 ~{\rm MHz}}$, corresponding to a loaded ${Q}$-factor ${\sim 1.8 \times 10^6}$. \textbf{(c)} Heterodyne beat signal between the sideband of soliton Kerr frequency comb and the narrow-linewidth reference laser. The measured beat signal is fitted with the Voigt profile (cf. Methods).  \textbf{(d, e)} Showcase of multiple dissipative solitons formed in ${\rm{Si_3N_4}}$ microresonators, in the breathing state \textbf{(d)} as well as in the low-noise stable soliton state \textbf{(e)}, the fitting of the spectrum envelope (green-solid lines) further shows the relative position of solitons circulating in the micro-ring cavity (Schematic insets). The low-frequency RF spectra corresponding to breather solitons are also shown as insets. Spectra in \textbf{(d)} and \textbf{(e)} are generated in ${\rm{Si_3N_4}}$ microresonators with a free spectral range (FSR) of ${\sim 88 ~{\rm GHz}}$ and ${\sim 92 ~{\rm GHz}}$, respectively.
  \label{fig_soliton}
}
\end{figure*} 

We next investigated if self-injection locking can also be observed in devices with an electronically detectable mode spacing ($ {149 ~{\rm GHz}}$, and ${<100 ~{\rm GHz}}$), and critically if it can also enable operation in a regime where DKS are formed concurrently.
Figure \ref{fig_switch}(a) shows the self-injection-locked Kerr comb generation in a microresonator with an FSR of ${149 ~{\rm GHz}}$. Significantly, not only were Kerr combs observed but also switching into the DKS regime \cite{herr2014temporal}.
Upon self-injection locking, and via current tuning we first excite a Kerr comb in a low-coherence state, as evidenced by the noise in the low-frequency RF spectrum [inset in Fig. \ref{fig_switch}(a)].
We emphasize that for such low repetition rates the amplitude noise is still a valid indicator of the frequency comb's coherence, in contrast to terahertz mode spacing resonators where the noise can be located at high RF frequencies (${>1 ~{\rm GHz}}$) \cite{pfeiffer2017octave}.
Importantly, we observe, that upon increasing the current to the diode further ( $\mathcal{O}$(mA)), which leads to a laser detuning increase by injection pulling, the low-coherence comb state is turned into an intermediate oscillatory state. That can be identified as a breather DKS [Fig. \ref{fig_switch}(b)] \cite{lucas2017breathing}, where the soliton exhibits periodic oscillations.
The RF spectrum shows the breathing frequency at ${\sim 490~{\rm MHz}}$  exhibiting harmonics, see inset in Fig. \ref{fig_switch}(b). Such soliton breathing dynamics, i.e. breather DKS, have been studied previously \cite{lucas2017breathing}, and in particular the breathing frequency depends on the laser detuning. The observation of a DKS breathing state demonstrates that the injection pulling enables operation in the effectively red detuned regime, required for soliton generation. 
Further increasing the laser current, we observe a transition to a low-noise comb state, demonstrating the formation of stable DKS as shown in  Fig. \ref{fig_switch}(c). The spectral envelope of the frequency comb exhibits a secant-squared profile, corresponding to a single soliton circulating in the resonator, with the breathing oscillations absent from the RF spectrum [inset in Fig. \ref{fig_switch}(c)]. This transition, which we induce here by current tuning only, has been achieved in previous work by tuning the laser over the resonance from the blue to the effectively red detuned side \cite{herr2014temporal}. Most significantly, to corroborate operation in the  soliton state we verify the coherence via a heterodyne beatnote measurement \cite{Cundiff2003comb}.
The heterodyne beatnote of a soliton comb tooth with a narrow linewidth reference laser is shown in Fig. \ref{fig_soliton}(c).
The measured heterodyne beatnote linewidth is comparable to that of the injection-locked laser [cf. Fig. \ref{fig_kickOn}(e)], i.e.  the Gaussian linewidth is ${201 ~{\rm kHz}}$ and the estimated short-time  Lorentzian linewidth (that describes the wings of the beatnote only) is ${1 ~{\rm kHz}}$. These values indicate no degradation of the coherence during the process of soliton comb generation via laser self-injection locking.

Moreover, DKS formation via laser self-injection locking was also observed in ${\rm{Si_3N_4}}$ microresonators with FSRs below ${100 ~{\rm GHz}}$, an electronically detectable repetition rate, where due to the high ${Q}$-factors {$({Q_0} \sim 8 \times 10^6)$} enabled by the photonic Damascene reflow process, soliton combs could still be generated \cite{liu2018ultralow-power}.
Figure \ref{fig_soliton}(a,b) show a dispersion measurement of the microresonator (cf. Methods), where the FSR is read as ${92.4 ~{\rm GHz}}$.
The parabolic dispersion profile shows quadratic contribution from an anomalous group velocity dispersion (GVD) to be: ${D_2 / 2\pi \approx 1.56 ~{\rm MHz}}$, centered at a wavelength ${\sim 1540 ~{\rm nm}}$.
The loaded resonance linewidth $\kappa / 2\pi$ is ca.  ${110 ~{\rm MHz}}$ [Fig. \ref{fig_soliton}(b)], corresponding to an over-coupled regime of the microresonator (the intrinsic loss rate is ${\kappa_0 / 2\pi< 30 ~{\rm MHz}}$). In these type of microresonators, multiple dissipative solitons are observed, shown in Fig. \ref{fig_soliton}(d,e), not only in the breathing state but in the low-noise stable soliton state as well.
The spectral envelope reveals a multi-soliton state as a result of interfering Fourier components of the solitons.
By fitting these spectral envelopes (cf. Methods), we can resolve the number of solitons and estimate their relative positions, illustrated as insets in Fig. \ref{fig_soliton}(d,e). The overall transmitted optical power, consisting of both the comb power and the residual pump power, is measured ${\sim 11 ~{\rm mW}}$ (cf. Methods).

In summary, we have demonstrated a route to an ultra-compact, cost-effective soliton frequency comb in photonic integrated ${\rm Si_3N_4}$ microresonators, via laser self-injection-locking with off-the-shelf laser diodes. We observed power-efficient soliton combs in  microresonators with different FSRs, particularly for FSR below ${100 ~{\rm GHz}}$.
This approach offers a dramatic reduction in size, cost and weight, and also offers simplified heterogeneous integration, in particular as no wafer bonding is required unlike for silicon photonic III-V lasers. This approach provides a route to scalable manufacturing of soliton microcombs for future high-volume applications.

\section*{Methods}
\label{Methods} 
\label{sec_method}
\section*{Photonic integrated ${\mathbf {Si_3N_4}}$ microresonator chip}
The photonic integrated ${\rm{Si_3N_4}}$ chips are fabricated by using the photonic Damascene reflow process. Waveguide and resonator patterns were defined by deep-UV stepper lithography and transferred to the ${\rm{SiO_2}}$ preform via dry etching. A preform reflow step was used to reduce the waveguide sidewall roughness caused by dry etching \cite{pfeiffer2016damascene,pfeiffer2018probing, pfeiffer2018damascene}, allowing for ultra-smooth waveguides and leading to high-${Q}$ factors for the microresonators. Optimized chemical mechanical polishing (CMP) allows precise control of the waveguide height to ${750 \pm 20 ~{\rm nm}}$, measured over the full 4-inch wafer scale. No top cladding was deposited onto the ${\rm{Si_3N_4}}$ waveguides. The precise dimension control by both the lithography (mainly in the waveguide width) and CMP (in the height) enables samples of the same design to have the identical geometry at different positions on the wafer.


The microresonator is coupled to the bus waveguide on the chip through the evanescent field.
Light is coupled onto the Si$_3$N$_4$ chip via double inverse nanotapers \cite{liu2018doubleinverse} on the bus waveguides at both the input and output facets, i.e. from the laser diode chip to the microresonator chip and from the microresonator chip to a lensed fiber which collects the comb spectrum.
In addition, the bus waveguide's geometry is designed to achieve a high coupling ideality with reduced parasitic losses \cite{pfeiffer2017coupling}.

The microresonator dispersion can be extracted by measuring the transmission spectrum, which is calibrated by a standard optical frequency comb \cite{del2009frequency, liu16dispersion}. The dispersion of the microresonator is represented in terms of resonant frequency deviation with respect to a linear grid, namely:
\begin{equation}
D_\mathrm{int} = \omega_{\mu} - ({\omega_0 + \mu D_1}) = \sum_{m \ge 2}{\frac{\mu^{m}D_{m}}{m!}}
\label{eqs_Dint}
\end{equation}
where ${\omega_{\mu}}$ is the physical resonant frequencies of the microresonator. A central resonance (to which the laser is injection locked) is given the index ${\mu = 0}$. ${D_1 = 2\pi \times {\rm FSR}}$ is the repetition frequency. The second order element ${D_2}$ is the group velocity dispersion (GVD) of the microresonator and ${D_2 > 0}$ represents the anomalous GVD. When the dispersion is described to the second order, the dissipative and nonlinear optical resonator can be described by the Lugiato-Lefever equation \cite{LLugiato-Lefever1987PRL}, which is equivalent to the coupled mode equation. Each resonance is fitted using the model based on coupled mode theory \cite{gorodetsky2000rayleigh, kippenberg2002coupling} from the transmission spectrum.
The resonance linewidth reflects the total loss rate (${\kappa}$) of the microresonator, which consists of both the intrinsic loss rate (${\kappa_0}$) and the external coupling rate ${\kappa_{\rm ex}}$, i.e. ${\kappa = \kappa_0 + \kappa_{\rm ex}}$.
To extract the intrinsic ${Q}$-factor (${Q_{\rm 0}}$), highly under-coupled microresonators are measured, i.e. ${\kappa_{\rm ex} \to 0}$.

In this work, there are three sets of ${\rm{Si_3N_4}}$ microresonators in terms of different FSRs: ${\sim 1 ~{\rm THz}}$, ${\sim 150 ~{\rm GHz}}$, and ${< 100 ~{\rm GHz}}$.
The microresonator corresponding to results shown in Fig. \ref{fig_kickOn} has: ${Q_{\rm 0} \approx 6 \times 10^6}$, ${{\rm FSR} = 1.02 ~{\rm THz}}$, ${D_2 / 2 \pi \approx 188 ~{\rm MHz}}$, for fundamental TE mode. The microresonator width is ${1.53 ~{\rm \mu m}}$.
The microresonator corresponding to results shown in Fig. \ref{fig_switch} has: ${Q_{\rm 0} \approx 6.5 \times 10^6}$, ${{\rm FSR} = 149 ~{\rm GHz}}$, ${D_2 / 2 \pi \approx 3.90 ~{\rm MHz}}$ (fundamental TE mode), the microresonatore width is ${1.58 ~{\rm \mu m}}$.
The microresonators corresponding to results shown in Fig. \ref{fig_soliton} have: ${Q_{\rm 0} \approx 8.2 \times 10^6}$, (for Fig. \ref{fig_soliton}(d)) ${{\rm FSR} = 88.6 ~{\rm GHz}}$, ${D_2 / 2 \pi \approx 1.10 ~{\rm MHz}}$ (fundamental TE mode), the microresonator width is ${1.58 ~{\rm \mu m}}$; (for Fig. \ref{fig_soliton}(e)) ${{\rm FSR} = 92.4 ~{\rm GHz}}$, ${D_2 / 2 \pi \approx 1.56 ~{\rm MHz}}$ (fundamental TE mode), the microresonator width is ${1.58 ~{\rm \mu m}}$. 

Such high ${Q}$-factors have already enabled direct soliton comb generation in microresonators without amplification of the seed laser \cite{liu2018ultralow-power}.
The threshold power for parametric oscillation can be as low as sub-milli-Watt (critical coupled), which is calculated as:
\begin{equation}
P_{\rm th} = \frac{\kappa^2 n^2 V_{\rm eff}}{4 \omega c n_2}
\end{equation}
where ${n}$ is the refractive index, ${V_{\rm eff}}$ indicates the effective modal volume, ${\omega}$ is the angular frequency of light, ${c}$ the speed of light in vacuum, and ${n_2}$ is the nonlinear refractive index.
For ${\rm{Si_3N_4}}$ microresonators with ${{\rm FSR} \sim 1 ~{\rm THz}}$, we have ${n \approx 1.9}$, ${V_{\rm eff} \approx 1.5 \times 10^{-16} ~{\mu m}^3}$, and ${n_2 \approx 2.4 \times 10^{-19} ~{\rm m^2/W}}$. Hence, the threshold power is as low as ${P_{\rm th} \approx 0.62 ~{\rm mW}}$.

\section*{Dissipative Kerr soliton comb power}
Multiple DKS in the microresonator with ${{\rm FSR} \sim 92.4 ~{\rm GHz}}$ are generated when applying a current ${\sim 280 ~{\rm mA}}$  to the diode chip, corresponding to an optical output power of ${\sim 50 ~{\rm mW}}$. 
The output power is measured as ${\sim 11 ~{\rm mW}}$, collected by using a lensed fiber at the output chip facet, indicating a coupling efficiency of ${\sim 22 \%}$ (overall insertion loss ${-6.6 ~{\rm dB}}$). 
The optical power in the bus waveguide is estimated to be  ${\sim 23.5 ~{\rm mW}}$, which has been demonstrated sufficient to excite DKS in high-${Q}$ ${\rm{Si_3N_4}}$ microresonators \cite{liu2018ultralow-power}.


\section*{Heterodyne beat signal and fitting function}
The heterodyne measurement is used to assess the coherence of the generated soliton comb, as its lineshape reveals the frequency noise spectral density with respect to the reference laser.
In fact, the frequency noise may consist of both the white noise (resulting in a Lorenztian lineshape) and the flicker noise (corresponding to a Gaussian lineshape).
Therefore, we employ the Voigt profile \cite{stephan2005laser} to fit the beat signal, which represents the convolution of the Lorenztian (${L(f)}$) and the Gaussian (${G(f)}$) lineshapes, i.e.: 
\begin{equation}
V(f) = \int_{- \infty}^{+ \infty } G(f';\sigma) L(f-f';,\psi) df',
\label{eq:voigt_profile}
\end{equation}
\begin{equation}
G(f;\sigma) = \frac{\exp^{-f^2/2 \sigma^2}}{\sigma \sqrt{2 \pi}},
\end{equation}
\begin{equation}
L(f;\psi) = \frac{\gamma}{\pi(f^2 + \psi^2)},
\end{equation}
where ${f}$ indicates the frequency shift with respect to the center of the beat signal, in the radio frequency domain, and ${\sigma}$ and ${\psi}$ scale the linewidth. 
To initiate the fitting we assume that, on the wings of the beat profile, the signal is mostly contributed by the white noise that determines the intantaneous linewidth described by ${\psi}$. 
In contrast, around the center of the beat profile, the signal is also contributed by flicker noise depending on e.g. the acquisition time of the ESA, as well as the stability of current or temperature controller. This part of noise is scaled by ${\sigma}$.
The full width at half maximum (FWHM) of the Gaussian lineshape is then ${\Delta f_{\rm G} = 2 \sigma}$ and ${\Delta f_{\rm L} = 2 \psi}$ for the Lorentzian.

\def \bwModeN {\Delta \mu}
\section*{Dissipative Kerr soliton comb spectral fitting}
It is known that ${N}$ identical solitons circulating in the resonator produce a spectral interference on the single soliton spectrum \cite{herr2014temporal, brasch2016photonic}:
\begin{equation}
S^{(N)}(\mu) = S^{(1)}(\mu) \left( N + 2 \sum_{j\neq l} \cos \Big(\mu(\phi_j-\phi_l)\Big) \right)
\label{eq:multisolSpectrum}
\end{equation}
Here ${\phi_i\in[0,2\pi]}$ is the position of the ${i}$-th pulse along the cavity roundtrip, ${\mu}$ is the comb mode index relative to the pump laser frequency and $S^{(1)}(\mu)$ is the spectral envelope of a single soliton following an approximate secant hyperbolic squared:
\begin{equation}
S^{(1)} \approx A \operatorname{sech}^2\left( \dfrac{\mu - \mu_c}{\bwModeN} \right)
\end{equation}
where $A$ is the power of the comb lines near the pump and $\bwModeN$ is the spectral width of the comb (in unit of comb lines) and $\mu_c $ is the central mode of the soliton (to account for soliton recoil or self frequency shift). Knowing the comb repetition rate $f_r$, the spectral width (or pulse duration) can be retrieved: $\Delta f = f_r \, \bwModeN$.
The spectral envelope of the single or multiple soliton states are fitted using the following procedure: First, the peaks $\tilde{S}(\mu)$ constituting the frequency comb are detected and labeled with their relative mode index from the pump $\mu$, and the pump mode is rejected. The number of solitons $N$ is estimated by taking the inverse Fourier transform of this spectrum, which yields the autocorrelation of the intracavity waveform, and detecting its peaks~\cite{brasch2016photonic}. The set of fitting parameters$\left\lbrace A,\bwModeN,\mu_c, \phi_i|i\in\llbracket 2,N\rrbracket  \right\rbrace$
 is defined accordingly (the position of one soliton is arbitraly set to zero) and the expression \eqref{eq:multisolSpectrum} is fitted to the experimental points $\tilde{S}(\mu)$. When $N$ solitons are perfectly equi-spaced, the repetition is multiplied by $N$ and the single soliton expression can be fitted on every $N$ line.

\section*{Funding Information}
 This publication was supported by Contract HR0011-15-C-0055 (DODOS) from the Defense Advanced Research Projects Agency (DARPA), Defense Sciences Office (DSO). 
This work was supported by funding from the Swiss National Science Foundation under grant agreement No. 163864 (Swiss-Russian JRP), Russian Science Foundation (RSF) (17-12-01413) and EU H2020 FET – OPTIMISM (grant no. 801352). H.G. acknowledges support by funding from the European Union’s Horizon 2020 research and innovation programme under Marie Sklodowska-Curie IF grant agreement No. 709249.
M.K. and E.L. acknowledge the support from  the European Space Technology Centre with ESA Contract No. 
4000116145/16/NL/MH/GM and 4000118777/16/NL/GM respectively.
All samples were fabricated in the Center of MicroNanoTechnology (CMi) at EPFL. We acknowledge Bahareh Ghadiani for the assistance in sample fabrication. We thank M.H. Anderson for useful discussion and assistance in manuscript preparation. 

\section*{Author contributions}
A.S.R., A.S.V., H.G., S.E.A., J.D.J., A.S.G., N.G.P. and M.K. setup the experiment and performed measurements. J.L. designed and fabricated the Si$_3$N$_4$ microresonator samples. A.S.R. and J.L. characterized and analyzed the Q and dispersion of the Si$_3$N$_4$ microresonator samples. A.S.R., H.G., A.S.V., A.E.S., J.L., E.L. and R.R.G. processed and analyzed the data. T.J.K., H.G., A.S.R., A.S.V., M.K. and J.L. wrote the manuscript with input from all authors. M.L.G. initiated the project and supervised the project from the Russian side. T.J.K. supervised the overall project.

\bibliographystyle{apsrev4-1}
\bibliography{biblography}

\end{document}